\documentclass[
    ,final            
  ]
  {aipproc}

\layoutstyle{6x9}

\begin{document}

\title{(Two) Open Questions in Stellar Nuclear Physics
\footnote{Work Supported by USDOE Grant No. DE-FG02-94ER40870.}}

\author{Moshe Gai}{
  address={Laboratory for Nuclear Science at Avery Point, \\
  University of Connecticut, 1084 Shennecossett Road, Groton, CT 06340.\\
  gai@uconn.edu, http://www.phys.uconn.edu}
}

\begin{abstract}
 No doubt, among the most exciting discoveries of the third millennium thus far 
 are {\bf Oscillations of Massive Neutrinos} and {\bf Dark Energy} that leads to an 
 accelerated expansion of the Universe.  Accordingly, Nuclear Physics 
 is presented with two extraordinary challenges: the need for precise (5\% or better) 
 prediction of solar neutrino fluxes within the Standard Solar Model, and the need 
 for an accurate (5\% or better) understanding of stellar evolution and in particular of 
 Type Ia super nova that are used as cosmological standard candle. In contrast, 
 much confusion is found in the field with contradicting data and strong statements 
 of accuracy that can not be supported by current data. We discuss an 
 experimental program to address these challenges and disagreements.
 
\end{abstract}

\maketitle


\section{Introduction}

During the last few years extraordinary discoveries in fundamental Physics were made  
using stellar objects as close as the sun, and as far away as the most distant 
Type Ia supernova (SNeIa) at the far end of the observed Universe. These discoveries have 
fundamentally altered our view of the observed universe and hint that yet several more discoveries
are soon to come. They were possible in part due to advances in Stellar Nuclear Physics, but 
they demand yet even higher precision of our knowledge of stars. The Standard Solar Model (SSM) 
\cite{Book} has been confirmed and the three decade persistent "solar neutrino problem" was solved 
by introducing neutrino oscillations \cite{SNO1}. The $^8B$ solar neutrino flux was measured 
with 7.3\% accuracy \cite{SNO2} and extracted from a global analysis of solar and reactor 
neutrino experiments \cite{Bah03} with 4\% accuracy. Type Ia supernova (SNeIa) were used as 
standard cosmological candles \cite{Phillips} and a recent acceleration in the expansion rate 
of the universe was suggested \cite{Perl}. The suggested accelerated expansion was confirmed 
by the WMAP experiment to arise from dark energy that constitute approximately 70\% of the 
observed universe \cite{WMAP}.

\section{The Standard Solar Model}

The Standard Solar Model is dependent on nuclear inputs and the most critical ones are 
cross sections of nuclear reactions \cite{Adel} at solar conditions of central temperature of 
15.7 MK and central density of approximately 150 $g/cm^3$. 
The two most important reaction cross sections that must 
be measured with an accuracy of 5\% or better are the $^7Be(p,\gamma)^8B$ reaction and the 
$^4He(^3He,\gamma)^7Be$ reaction and the corresponding $S_{17}(0)$ and $S_{34}(0)$ defined 
in Ref. \cite{Adel}. 

A major effort on measuring $S_{17}(0)$ with high accuracy was carried out in several 
labs and agreement among high precision data collected at GSI \cite{Sch03}, Weizmann 
\cite{Weiz} and Seattle \cite{Seatt} was found. Most amazing is the excellent agreement 
between the Weizmann data that were measured with a $^7Be$ target and the GSI data that employed 
the Coulomb dissociation method. However as shown in Fig. 1 the slopes of these three results 
are sufficiently different. The d-wave correction to $S_{17}(0)$ on the other hand is 
directly related to this slope, and thus it is ill determined. Since the d-wave correction 
reduces $S_{17}(0)$ by as much as 15\%, 
it precludes an accurate extrapolation of $S_{17}(0)$. This 
conclusion contradicts the strong statement of the Seattle group \cite{Seatt} that 
$S_{17}(0)$ has been determined with a theoretical uncertainty of 2.5\%. 
This issue must be resolved by future high precision measurements of the slope, most likely 
with $^7Be$ beams \cite{ISOLDE}, so as to allow accurate (5\% or better) 
extrapolation of $S_{17}(0)$. 

In contrast to the intensive work on $S_{17}(0)$, 
no progress what-so-ever was achieved on measuring $S_{34}(0)$ with 
high precision, and it is still poorly known with an error of 9\% \cite{Adel}. This inadequate 
situation must be improved in the near future as we expect the direct detection of 
$^7Be$ solar neutrinos. These measurements will conclude a four decade long quest by 
Nuclear Physicists for the nuclear inputs to the SSM. When the controversy on the composition 
of sun (Z/X) will also be resolved \cite{Basu}, it will allow high precision prediction 
of all solar neutrino fluxes including the $^8B$ neutrino flux. The high precision on one hand 
may provide a strong evidence for the SSM, but may also allow for a study of fundamental 
neutrino processes including oscillation to sterile neutrinos.

 \begin{figure}
  \includegraphics[height=.3\textheight]{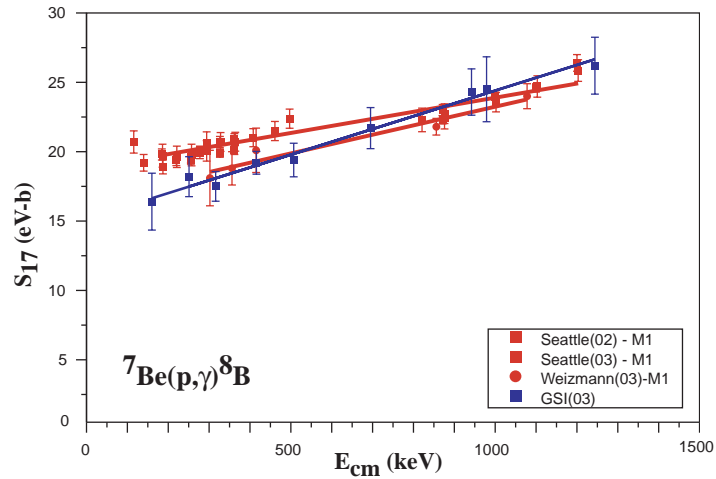}
  \caption{A comparison of the GSI \cite{Sch03}, Weizmann \cite{Weiz} and Seattle 
  \cite{Seatt} measurements of the astrophysical 
  cross section factor $S_{17}$ of the $^7Be(p,\gamma)^8B$  reaction, as defined in 
  \cite{Adel} and discussed in the text.}
\end{figure}

\section{Helium Burning and the C/O Ratio}

The C/O ratio at the end of helium burning is still poorly known, twenty years after 
it was declared by Willie Fowler the "holy grail" of Nuclear Astro-Physics \cite{Fowler}. 
This parameter is essential for almost all aspects of stellar evolution of massive stars, 
and most recently it was also suggested to be essential for understanding the light curve of 
SNeIa \cite{Hoef}. The finding of Hoeflich were recently challenged \cite{Hild}, 
but the C/O ratio is most certain to play a major role in our understanding of the Phillips 
empirical relationship of peak luminosity and the shape of the light curve of 
Type Ia supernova \cite{Phillips}. Since the Phillips relationship is at the very foundation of 
using SNeIa as standard cosmological candle it is essential to understand it. The new 
generation of dedicated space telescopes that will solely measure Type Ia supernova makes
it very important to understand SNeIa.

In order to measure the C/O ratio at the end of helium burning the cross section 
of the $^{12}C(\alpha,\gamma)^{16}O$ needs to be known at approximately 
300 keV, but thus far it was measured only down to approximately 1.2 MeV. The extrapolation of 
this cross section to stellar energies (300 keV) is particularly difficult due to the substantial 
contribution from bound states.  The properties of the bound states and their interference 
with quasi-bound states were thus far determined with the use of R-matrix theory. However, it 
now appears that the claimed accuracy of the R-matrix fits can not be substantiated. While the TRIUMF 
group quote an E1 astrophysical cross section factor with 25\% uncertainty \cite{TRIUMF}, Hale 
extracts a value that is eight times smaller \cite{Hale}. Similarly elastic scattering data was 
used by the Notre Dame group to extract the E2 S-factor with the claimed 20\% accuracy \cite{ND}. 
But this analysis in of itself was criticized for lack of theoretical foundation \cite{JM}, and 
the result turned out to be a factor of 2.5 smaller than extracted by the Stuttgart group 
\cite{Stutt} that used R-matrix theory to extrapolate angular distribution data of the 
$^{12}C(\alpha,\gamma)^{16}O$ capture reaction itself.

A most promising new approach to measure both the E1 and E2 astrophysical cross section factors 
of the $^{12}C(\alpha,\gamma)^{16}O$ reaction at energies as low as 700 keV  emerged with 
the use of the High Intensity $\gamma$amma Source (HI$\gamma$S) at the TUNL lab at Duke
\cite{HIGS}. In this experiment one will study the 
photodisintegration of $^{16}O$ with an Optical Readout 
Time Projection Chamber (TPC). The anticipated results of the HI$\gamma$S facility are shown 
in Fig. 2, as compared to the disagreeing results discussed above.

 \begin{figure}
  \includegraphics[height=.3\textheight]{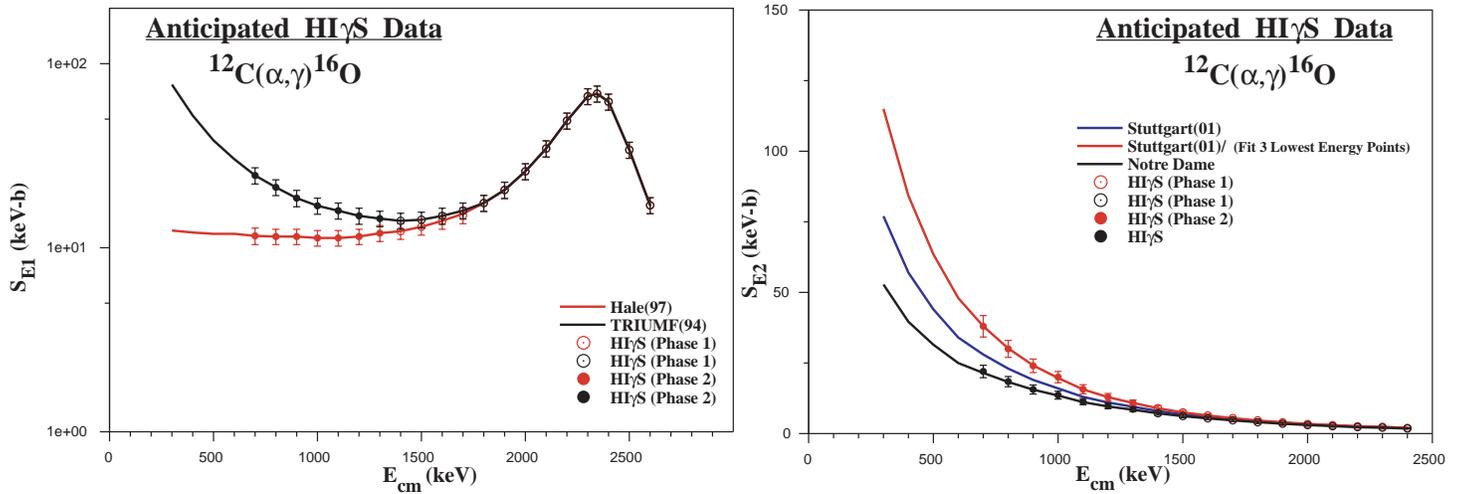}
  \caption{Anticipated HI$\gamma$S data on the E1 S-factor as compared to the values  
  quoted by the TRIUMF collaboration \cite{TRIUMF} and Hale \cite{Hale}, and on the 
  E2 S-factor as compared to the results of the Notre Dame \cite{ND} and the Stuttgart 
  groups \cite{Stutt}.}
  
\end{figure}




\vfill\eject

\end{document}